\documentclass[aps,reprint,groupedaddress,showpacs,floatfix,superscriptaddress]{revtex4-2}
\usepackage[utf8]{inputenc}
\usepackage[pdftex]{graphicx} 
\usepackage{bm} 
\usepackage{amssymb} 
\usepackage{amsmath} 
\usepackage{csquotes}
\usepackage{float}
\usepackage{gensymb}
\usepackage{chemformula}
\usepackage{orcidlink}
\usepackage{hyperref}

\begin{document}
\title{Electronic excited states in deep variational Monte Carlo}
\author{M. T. Entwistle\orcidlink{0000-0001-8049-8563}}
\thanks{M. T. Entwistle and Z. Schätzle contributed equally to this work.}
\affiliation{FU Berlin, Department of Mathematics and Computer Science, Arnimallee 12, 14195 Berlin, Germany}
\author{Z. Schätzle\orcidlink{0000-0002-5345-6592}}
\thanks{M. T. Entwistle and Z. Schätzle contributed equally to this work.}
\affiliation{FU Berlin, Department of Mathematics and Computer Science, Arnimallee 12, 14195 Berlin, Germany}
\author{P. A. Erdman\orcidlink{0000-0003-4626-2869}}
\affiliation{FU Berlin, Department of Mathematics and Computer Science, Arnimallee 12, 14195 Berlin, Germany}
\author{J. Hermann\orcidlink{0000-0002-2779-0749}}
\email{science@jan.hermann.name}
\affiliation{FU Berlin, Department of Mathematics and Computer Science, Arnimallee 12, 14195 Berlin, Germany}
\author{F. Noé\orcidlink{0000-0003-4169-9324}}
\email{franknoe@microsoft.com}
\affiliation{Microsoft Research AI4Science}
\affiliation{FU Berlin, Department of Mathematics and Computer Science, Arnimallee 12, 14195 Berlin, Germany}
\affiliation{FU Berlin, Department of Physics, Arnimallee 14, 14195 Berlin, Germany}
\affiliation{Rice University, Department of Chemistry, Houston, Texas 77005, USA}

\date{\today}

\begin{abstract}   
Obtaining accurate ground and low-lying excited states of electronic systems is crucial in a multitude of important applications. One \textit{ab initio} method for solving the Schrödinger equation that scales favorably for large systems is variational quantum Monte Carlo (QMC). The recently introduced deep QMC approach uses ansatzes represented by deep neural networks and generates nearly exact ground-state solutions for molecules containing up to a few dozen electrons, with the potential to scale to much larger systems where other highly accurate methods are not feasible. In this paper, we extend one such ansatz (PauliNet) to compute electronic excited states. We demonstrate our method on various small atoms and molecules and consistently achieve high accuracy for low-lying states. To highlight the method's potential, we compute the first excited state of the much larger benzene molecule, as well as the conical intersection of ethylene, with PauliNet matching results of more expensive high-level methods.
\end{abstract}

\maketitle

\section{Introduction}
The fundamental challenge of quantum chemistry, solid-state physics and many areas of computational materials science is to obtain solutions to the electronic Schrödinger equation for a given system, which in principle provide complete access to its chemical properties. The ground and low-lying excited states typically determine the behavior of a system and are therefore of the most interest in many applications. Understanding and being able to describe excited-state processes \cite{Lindh2020}, including a wide variety of important spectroscopy methods such as fluorescence, photoionization and optical absorption of molecules and solids, is key to the successful design of new materials.

Unfortunately, the Schrödinger equation cannot be solved exactly except in the simplest cases, such as one-dimensional toy systems or a single hydrogen atom. Accordingly, many approximate numerical methods have been developed which provide solutions at varying degrees of accuracy. Time-dependent density functional theory \cite{Kohn1965, Runge1984} (TDDFT) is the most popular method due to its computational efficiency, but has well known limitations \cite{Elliott2012, Fuks2015, Suzuki2017, Singh2019, Maitra2017, Ullrich2016}. Higher-accuracy methods have a computational cost that scales rapidly with system size --- the well established full configuration interaction \cite{Szalay2012} (FCI) and coupled cluster \cite{Sneskov2011} (CC) techniques scale $\sim \mathcal{O}(\text{exp}(N))$ \footnote{FCI scales exponentially, while truncated CI scales polynomially.} and $\sim \mathcal{O}(N^{5-10})$ \footnote{The scaling of CC depends on the particular method used: CC2 $\mathcal{O}(N^{5})$, CCSD $\mathcal{O}(N^{6})$, CCSD(T) $\mathcal{O}(N^{7})$, CC3 $\mathcal{O}(N^{7})$, CCSDT $\mathcal{O}(N^{8})$, CCSDT(Q) $\mathcal{O}(N^{9})$, CCSDTQ $\mathcal{O}(N^{10})$.} respectively, where $N$ is the number of electrons, thereby severely limiting their practical use. There is thus a huge need for \textit{ab initio} methods that scale more favorably with system size, allowing the modeling of practically relevant molecules and  materials.

Quantum Monte Carlo (QMC) techniques offer a route forward with their favorable scaling ($\mathcal{O}( N^{3-4})$) and therefore dominate high-accuracy calculations where other methods are too expensive \cite{Foulkes2001, Williams2020}. A state-of-the-art QMC calculation typically involves the construction of a multi-determinant baseline wavefunction through standard electronic-structure methods, which is augmented with a Jastrow factor to efficiently incorporate electron correlation, and then optimized through variational QMC (VMC) to obtain a trial wavefunction. This is then used within fixed-node diffusion QMC (DMC) to obtain a final electronic energy. The fixed-node approximation is used to avoid exponential scaling, with the drawback that the nodal surface of the trial wavefunction cannot be modified, which limits the accuracy of the DMC result \cite{Morales2012}. A more expressive baseline wavefunction can improve upon this but traditional DMC often needs thousands to hundreds of thousands of determinants to reach convergence \cite{Benali2020}. Additionally, DMC only provides the final energy, restricting the calculation of other electronic properties \cite{Austin2012}. Both of these limitations can, in principle, be resolved at the VMC level, with the accuracy of VMC constrained only by the flexibility of the trainable wavefunction ansatz. So far, these techniques have mostly been developed for ground-state calculations, with different extensions proposed to address excited states \cite{Foulkes2001, Ceperley1988, Blunt2015, Send2011, Dash2019, PinedaFlores2019, Zhao2016, Shea2017, Blunt2019, Choo2018, Pathak2021}.

Recently, the new \textit{ab initio} approach of deep VMC methods has been introduced \cite{Hermann2020,Pfau2020,Choo2020,HanZhangE2019} and subsequently further extended and improved \cite{Spencer2021,Scherbela2021,Gao2021}. In particular, PauliNet \cite{Hermann2020} and FermiNet \cite{Pfau2020} were the first methods to demonstrate that highly accurate ground-state results for molecules could be obtained using deep VMC with a lower computational complexity and using orders of magnitude fewer Slater determinants typically employed in other methods that achieve similar accuracy.

In the same spirit as Carleo and Troyer proposed for optimizing quantum states in lattice models \cite{Carleo2017}, VMC is used in order to train a neural network model that represents the many-body wavefunction in an unsupervised fashion, i.e. in contrast to other quantum machine learning approaches the only input to the method is the Hamiltonian, and training data are generated on the fly by sampling from the current wavefunction model and minimizing the variational energy. In both PauliNet and FermiNet deep antisymmetric neural networks are used to represent the fermionic wavefunction in the real space of electron coordinates.

Recently, there has been much interest in developing deep learning methods for excited states \cite{Westermayr2021}. In this paper, we extend PauliNet towards the \textit{ab initio} computation of electronic excited states (see the \enquote{Methods} section for details). The input is again only the Hamiltonian of the quantum system. By employing a simple energy minimization and numerical orthogonalization procedure, we are able to obtain the lowest excited-state wavefunctions of a given system. The excited-state optimization makes use of a penalty method that minimizes the overlap between the $n$-th excited state and the lower-lying states in the spectrum. Optimization methods that introduce additional constraints have been used in the context of VMC before \cite{Pathak2021} and provide a simple way to obtain orthogonal states without explicit enforcement in the wavefunction ansatzes. Combining these techniques with the expressiveness of neural network ansatzes yields highly accurate approximations to excited states with direct access to the wavefunctions for the evaluation of electronic observables. Neural network-based methods have targeted low-lying excited states of one-dimensional lattice models \cite{Choo2018}, but have not been applied to first-principles systems.

We demonstrate our method on a variety of small- and medium-sized molecules, where we consistently achieve highly accurate total energies, outperforming traditional quantum chemistry methods. We also compute excitation energies, transition dipole moments and oscillator strengths, the main ground-to-excited transition properties, with the latter two known to be more sensitive to errors in the underlying wavefunctions than energies. In all test systems, we find PauliNet closely matches high-order CC and experimental results. Next, we show that our method can be applied in a straightforward manner to much larger molecules, using the example of benzene where we match significantly more expensive high-level electronic-structure methods. Finally, we demonstrate that PauliNet can be used to compute excited-state potential energy surfaces by modeling an avoided crossing and conical intersection of ethylene, a highly multi-referential problem.

\newcommand\T{\rule{0pt}{3.2ex}}       
\newcommand\B{\rule[-1.8ex]{0pt}{0pt}} 

\section{Results}
\subsection{Nearly exact solutions for small atoms and molecules}\label{sec:small_systems}
To demonstrate our method we start by applying it to a range of small atoms and molecules. We optimize the lowest-lying excited states and compute their vertical excitation energies for the ground-state equilibrium geometry (see Supplementary Information Table I), with each PauliNet wavefunction containing a maximum of 10 determinants. In all systems, we obtain highly accurate total energies and estimates of the first few excitation energies competitive with high-accuracy quantum chemistry methods.

\begin{figure}[h]
    \centering
    \includegraphics[width=\columnwidth]{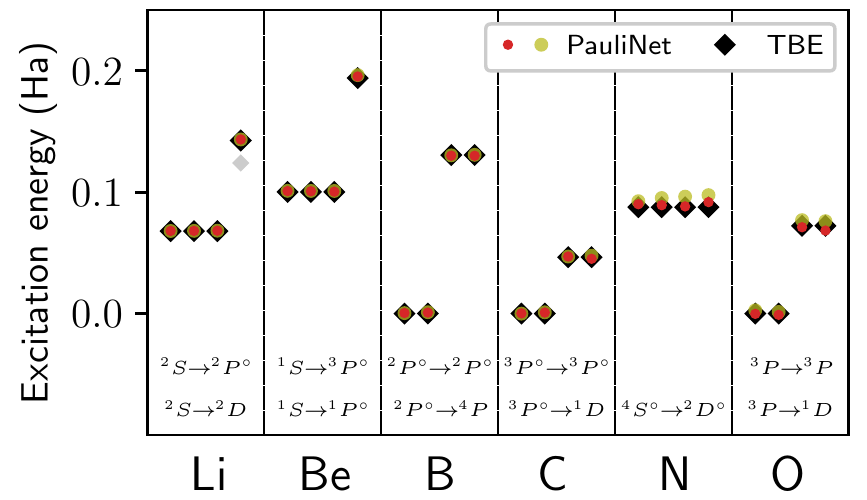}
    \caption{\textbf{Deep VMC obtains highly accurate excited states for single elements.} PauliNet results for the excitation energies (with (red) and without (yellow) variance matching (see the \enquote{Methods} section for details)) are compared to the theoretical best estimates (TBE) taken from the NIST database \cite{NIST}. Multiple PauliNet ansatzes with identical energies correspond to orthogonal degenerate states. For the TBE we have depicted four excitations per atom, taking account of the degeneracies. For all atoms we find the first excited state with high accuracy. For B, C and O the ground state is threefold degenerate. For these systems we choose one of the three states to compute excitation energies, resulting in transitions with a relative energy of zero. For Li and Be a further excitation energy is found. While we obtain the second excited state for Be, in Li we miss out intermediate states and instead find the transition from the ground state to the $^2D$ state. This can be related to the generic CASSCF initialization of the ansatzes. (The numerical data can be found in Supplementary Information Table II.)}
    \label{fig:atoms}
\end{figure}

\begin{figure*}[t!]
    \centering
    \includegraphics[width=\textwidth]{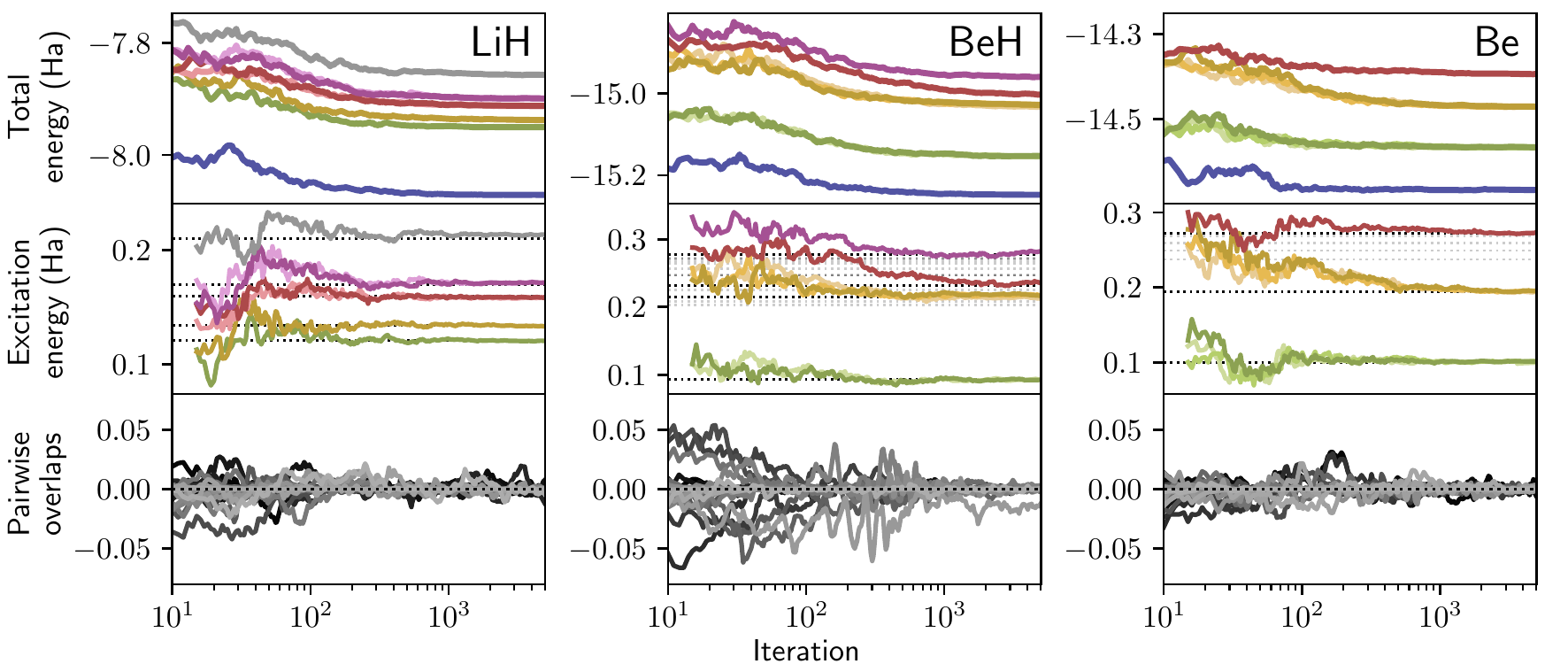}
    \caption{\textbf{Optimizing low-lying excited states for small molecules.} Several excited states of LiH, BeH and Be are approximated. The convergence of the total energies (upper row), excitation energies (middle row) and the pairwise overlaps between the wavefunctions (bottom row) are shown. For degenerate states multiple ansatzes attain the same energy. Dotted horizontal lines are excitation energies from FCI calculations and other highly accurate references \cite{bandeLiHPotentialEnergy2010, jasikElectronicStructureTimedependent2017, pitarch-ruizFullConfigurationInteraction2008, NIST}. Due to the initialization from the CASSCF baseline the wavefunctions start with a small overlap, which is retained throughout the optimization. (The numerical data can be found in Supplementary Information Table II.)
    }
    \label{fig:small-systems}
\end{figure*}

In Fig.~\ref{fig:atoms} the excitation energies of the lowest states are shown for several atoms. For all the atoms the excitation energies are obtained within 4~mHa of the theoretical best estimates (TBE) \cite{NIST}. Due to the high degree of symmetry the atoms exhibit degeneracies, that is, multiple orthogonal states can be found with the same energy. Being subject to the orthogonalization constraint, PauliNet approximates all orthogonal states of an energy level individually, which is observed by attaining multiple results at the same energy level. The multiplicity of the exact solution can be obtained theoretically by considering the electronic configurations of the atoms and is reproduced within our experiments.

We then compute a larger number of excited states for LiH, BeH and Be. In each experiment, we optimize eight ansatzes in parallel. In Fig.~\ref{fig:small-systems} we illustrate the training process by plotting the convergence of the total energies and excitation energies. Additionally, we plot the training estimates of the pairwise overlaps of the wavefunctions, which remain small throughout the optimization process. We confirm that the final overlaps are near-zero by exhaustively sampling the trained wavefunctions, thereby obtaining well-converged Monte Carlo estimates (see Supplementary Information Table VI). Based on the degeneracies we find a total of five (LiH), four (BeH) and three (Be) distinct excitation energies, respectively. The excitation energies match those from reference values, and in particular, we find that for all systems studied here we reliably obtain the first excited state, and apart from one case also the second excited state. However, especially for clusters of higher-lying excited states with similar energies, we typically do not find all members of the cluster. In these cases, which states are found depends on the initialization of our ansatzes, as well as the total number of states that are being sought. To give a transparent picture of the capabilities of our method, in this work we have refrained from optimizing the CASSCF baseline in order to find all possible excitations.

\subsection{Highly accurate wavefunctions: transition dipole moments and oscillator strengths}
Total energies and vertical excitation energies are the primary focus when benchmarking excited-state methods as they are readily available from many theoretical models and provide a good initial guess of a particular method's accuracy. However, they provide only a partial characterization of the electronic states, and while a method in question may give accurate energies, other quantities of key importance may be inaccurate \cite{Bremond2018, Tajti2019, Tajti2020}.

Transition dipole moments (TDM) and oscillator strengths are two principal ground-to-excited transition properties and are of great interest. TDMs determine how polarized electromagnetic radiation will interact with a system due to its distribution of charge, and therefore determine transition rates and probabilities of induced state changes. In the electric dipole approximation, the TDM between two states $i$ and $j$ is given by
\begin{equation}
    \mathbf d_{ij} = \langle \psi_{i}|\hat{\boldsymbol \mu}| \psi_{j}\rangle,
\end{equation}
where $\hat{\boldsymbol\mu} = \sum_{k} q \hat{\mathbf r}_k$ is the sum over the position operator of each particle weighted by its charge, with $q=-e$ for electronic systems. We obtain the expectation value by Monte Carlo sampling according to Eq.~\eqref{eq:mixed_operator}. While the TDM is important for understanding a number of processes, including optical spectra, it is a complex vector quantity and not an experimental observable by itself. The closely related oscillator strength is what is inferred through experiment and is given by
\begin{equation}
    f_{ij} = \frac{2}{3} \Delta E d_{ij}^2,
\end{equation}
where $\Delta E$ is the excitation energy between states $i$ and $j$, and $d_{ij}^2$ is the dipole strength. It is known that, in addition to being more basis-set sensitive, $d_{ij}$ and $f_{ij}$ are both highly dependent on the quality of the trial wavefunctions \cite{Crossley1984} and represent a more rigorous test for \textit{ab initio} methods than just energies. 

Recently, transition energies and oscillator strengths for a variety of small molecules have been computed using high-order CC calculations, systematically extrapolating to the complete basis set (CBS) limit, and comparing to experimental results where possible, in order to supply a comprehensive set of theoretical benchmarks \cite{Loos2018, Loos2021}. In that spirit, we now use these results to benchmark the accuracy of oscillator strengths computed using PauliNet. Furthermore, we also compare to multi-reference CC (MR-CC) results where possible \cite{Bhattacharya2013}. We compute the first few electronic states for five molecules (BH, CH$^{+}$, H$_{2}$O, NH$_{3}$, CO), such that we obtain the first non-zero oscillator strength (within the dipole approximation) for each. All calculations \footnote{CH$^{+}$ was not included in the CC calculations in refs.~\onlinecite{Loos2018, Loos2021}. We instead compare to (MR-)CC results in ref.~\onlinecite{Bhattacharya2013}, using the same ground-state equilibrium geometry, which was obtained in a split-valence basis augmented with diffuse and polarization functions. See refs.~\onlinecite{Bhattacharya2013,Olsen1989} for more details.} are performed at the same ground-state equilibrium geometries as refs.~\onlinecite{Loos2018,Loos2021} (see Supplementary Information Table I) and using the same number of determinants ($\leq$ 10) as in Section ~\ref{sec:small_systems}. 

\begin{figure}
    \centering
    \includegraphics[width=0.95\columnwidth]{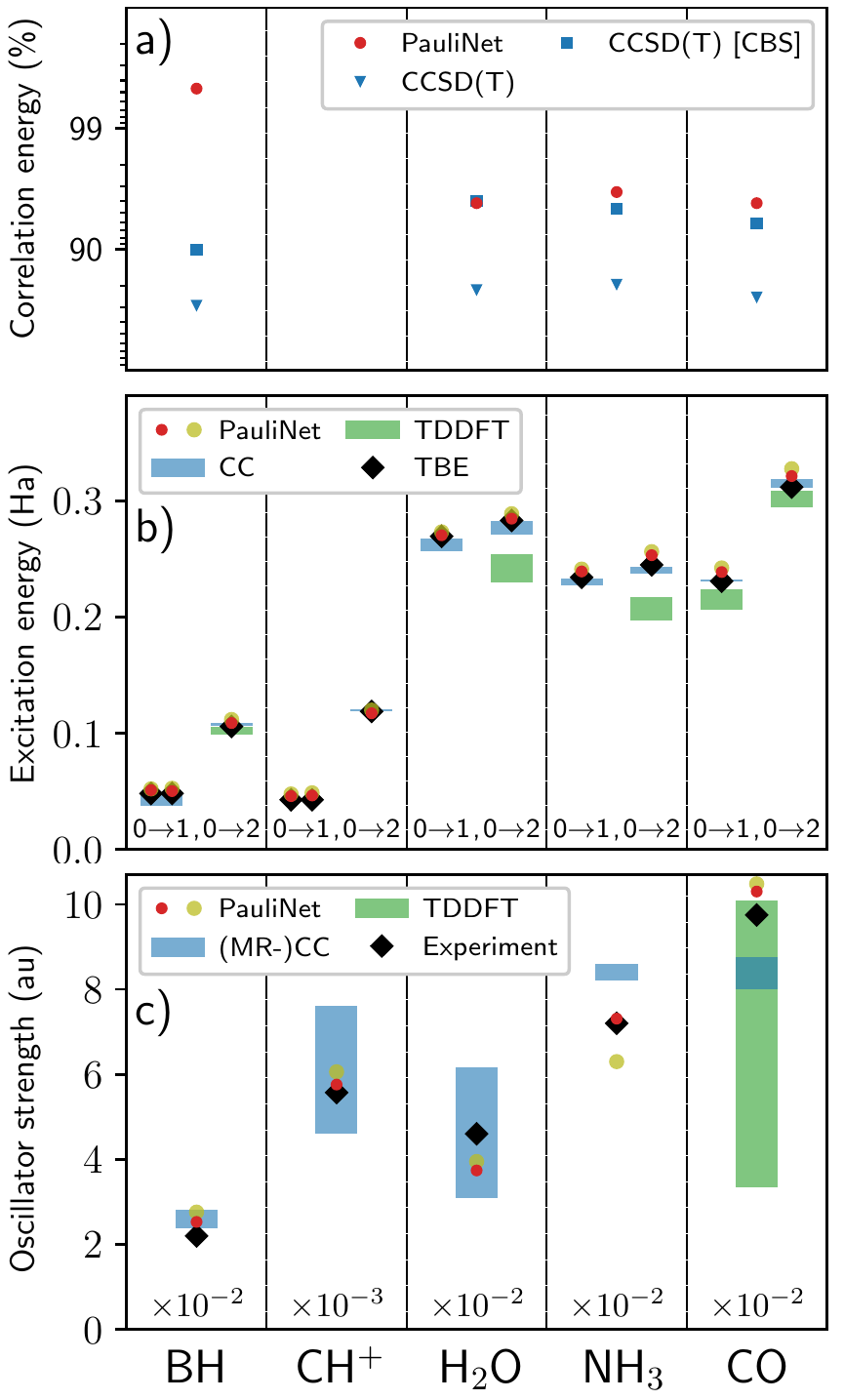}
    \caption{\textbf{Deep VMC obtains highly accurate excited-state energies and wavefunctions for small molecules.} \textbf{a} PauliNet recovers the same amount of correlation energy as high-order CC methods \cite{NIST}. (CH$^{+}$: No better reference energy to compare with.) \textbf{b} Lowest triplet (0$\rightarrow$1) and singlet (0$\rightarrow$2) excitation energies obtained using PauliNet (with (red) and without (yellow) variance matching), CC and TDDFT, with the TBE given. (BH and CH$^{+}$ exhibit degeneracy for the triplet state; CC is CCSD or higher, except for the triplet state of BH which includes CC2.) \textbf{c} Oscillator strengths computed for the 0$\rightarrow$2 transitions. PauliNet compares well to experiment in all systems and matches the accuracy of (MR-)CC results, demonstrating the quality of few-determinant PauliNet wavefunctions. (We have omitted a factor of two linked to degeneracy in BH and CO.) Refs: exact correlation energies \cite{Darragh2005, Giner2021, NIST}; excitation energies from CC \cite{Loos2018, Loos2021, Larsen2001, Bhattacharya2013, Kowalski2001, Kowalski2001_2, Cronstrand2004, Pawel2005}, TDDFT \cite{Fenglai2010, Cronstrand2004, Pawel2005, Carlo1999} and TBE \cite{Loos2018, Loos2021, Larsen2001, Biglari2014, Olsen1989}; oscillator strengths from (MR-)CC \cite{Loos2018, Loos2021, Bhattacharya2013, Barysz1995, Lane2008}, TDDFT \cite{Tawada2004} and experiment \cite{Douglass1989, Mahan1981, Thorn2007, Chen2019, Kang2015}. (The numerical data can be found in Supplementary Information Table III.)}
    \label{fig:os}
\end{figure}

Our results for all systems are shown in Fig~\ref{fig:os}. First, we compute the amount of correlation energy recovered in the ground state, and find PauliNet matches high-order CC methods (panel a). Second, we compute the excitation energy for each transition and find this to be close to the TBE, on par with CC and much more consistent than TDDFT where the accuracy depends on the molecule and on the exact TDDFT method used (panel b). Finally, we compare the oscillator strengths (for the $0\rightarrow2$ transition) in panel c. Even high-order methods such as CC and MR-CC can produce a spectrum of results depending on the expansion and basis set used, with this exacerbated in cheaper methods such as TDDFT (see the example of CO). In all systems, PauliNet compares well with experimental results, demonstrating the quality of deep VMC wavefunctions with just a minimal number of determinants.

\subsection{Application to larger molecules}\label{sec:benzene}
\begin{figure}
    \centering   
    \includegraphics[width=\columnwidth]{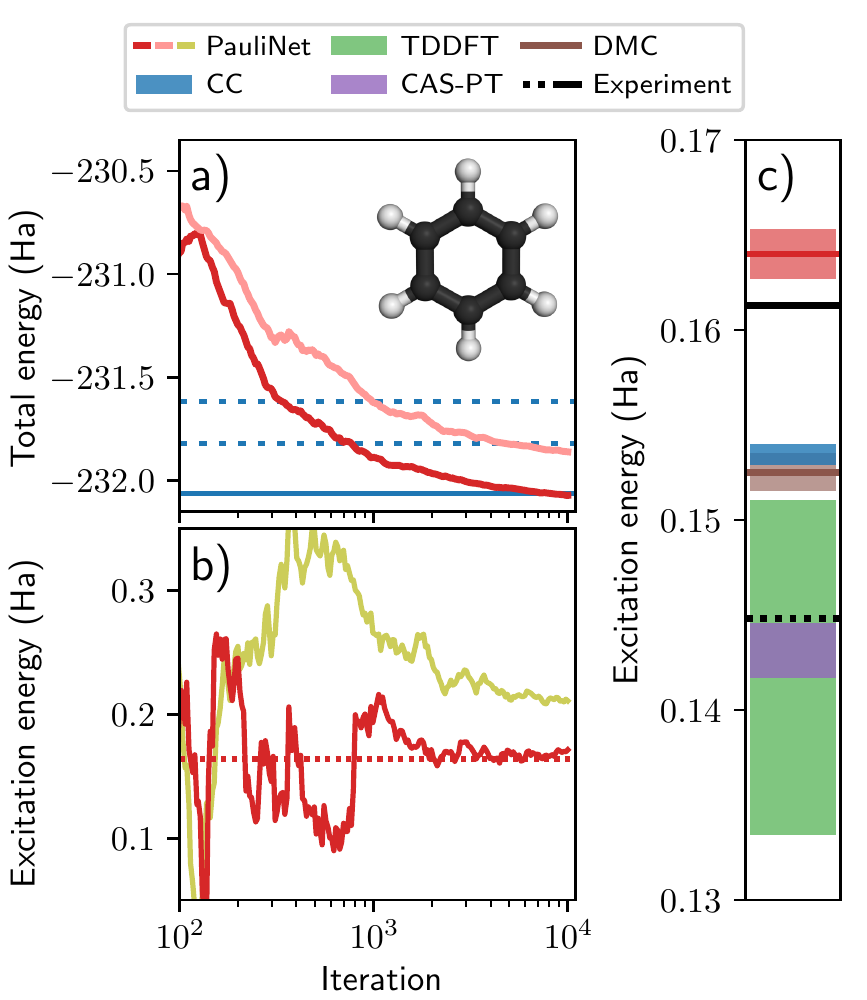}
    \caption{\textbf{Calculating the two lowest electronic states of the benzene molecule.} Inset: Benzene structure. \\ \textbf{a} Convergence of the total energies of the ground state (red) and excited state (light red) with training. Total energies of the ground state from CCSD(T) in the frozen-core approximation with the aug-cc-pV\textit{n}Z basis set ($n=$ D, T) (dashed blue), and full CCSD(T) at the CBS limit (solid blue) are shown \cite{NIST}. \textbf{b} Convergence of the excitation energy with training (with (red) and without (yellow) variance matching). \textbf{c} Excitation energy computed using PauliNet, TDDFT \cite{Carlo1999}, CC \cite{Loos2020}, DMC \cite{Pathak2021}, CAS-PT (\cite{Lorentzon1995} and calculations in openMolcas \cite{Fdez2019}) and Experiment \cite{Doering1969, Pathak2021} (adiabatic (dashed black) and vertical (solid black) excitation energies). (The numerical data can be found in Supplementary Information Table IV.)}
    \label{fig:Benzene}
\end{figure}

The previous two sections showed that we achieve highly accurate results across a range of small systems. While this is encouraging, traditional high-accuracy methods that are better established are readily available for such small systems. In this section, to demonstrate the potential of excited PauliNet, we show that it can be applied in a straightforward manner to significantly larger molecules. For this objective, we choose the example of the benzene molecule (panel a of Fig.~\ref{fig:Benzene}). Studies of its electronic structure and other properties are plentiful due to its importance in bio and organic chemistry, and with 42 electrons all-electron calculations will be extremely demanding or even intractable for a high-level description of its electronic states, depending on the theory level used.

Using a PauliNet ansatz with just 10 determinants, the same as in the much smaller systems, and slightly deeper neural networks (see Supplementary Information Table VII) we obtain very good total energies for the ground state and first excited state (upper left of Fig.~\ref{fig:Benzene}). We note the better accuracy than high-level CC calculations, with this signifying highly accurate wavefunctions that can be used to compute other observables, as demonstrated in the previous section. The computed excitation energy is also shown (right of Fig.~\ref{fig:Benzene}), with PauliNet compared against several experimental and theoretical results. The lower experimental result \cite{Doering1969} (dashed black line) quantifies an adiabatic excitation energy, i.e. the energy difference between the ground state and the excited state at the corresponding relaxed geometries. This quantity is corrected to obtain the vertical excitation energy \cite{Pathak2021} (solid black line), which omits nuclear relaxation and vibrational effects. As our calculations are performed at the ground-state equilibrium geometry, we are targeting the vertical excitation energy, and therefore consider this corrected experimental result to be closer to the ground truth. We find this to be slightly underestimated by high-order methods (CC, DMC), and slightly overestimated by PauliNet. In other systems (panel b of Fig~\ref{fig:os}) we notice a similar trend when comparing to the TBE.

PauliNet formally scales as $\mathcal{O}(N^4)$ with the number of electrons $N$, and in practice we observe a scaling behavior $\mathcal{O}(N^3)$ for the systems investigated so far, which is related to quadratic scaling of the neural network with an extra factor from the evaluation of the local energy. As PauliNet is currently implemented in a research code, which is not optimized for production purposes, the computational time will have a large prefactor which makes it computationally unfavorable to, e.g. CC methods for small molecules. However, its very favorable scaling in $N$ compared to $\mathcal{O}(N^{5-10})$ of high-level electronic-structure methods dominates for larger molecules, and this is clearly visible in benzene. For instance, ref.~\onlinecite{Eriksen2020} used several state-of-the-art methods to obtain accurate benzene ground-state energies, with calculations run on several CPU types in a highly parallel manner (see Supporting Information of ref.~\onlinecite{Eriksen2020} for details). PauliNet was run on a single RTX 3090 GPU at a fraction of the number of node hours. Although PauliNet is the computationally cheapest method in this comparison, it provides a significantly better (variational) ground-state energy than all methods ($\sim 0.48$ Ha lower). As all methods compared in Fig.~\ref{fig:Benzene} provide similar excitation energies, these cannot be used to group the methods into more or less accurate, but overall this data indicates that PauliNet and deep VMC methods in general have a very favorable cost/accuracy trade-off for molecules of the size of benzene and beyond.

\subsection{Multi-reference application: conical intersections}\label{sec:ethylene}
Molecular configurations that produce electronic states with similar energies are fundamental in photochemical applications. Such configurations can lead to several states mixing, meaning they are all necessary for an accurate description of a particular process. Conical intersections are produced when two states become degenerate and require the computation of excited-state potential energy surfaces. The modeling of energy surfaces near degeneracies is inherently multi-reference with significant electronic correlation and is thus a challenging application for electronic-structure methods. 

\begin{figure}
    \centering
    \includegraphics[width=\columnwidth]{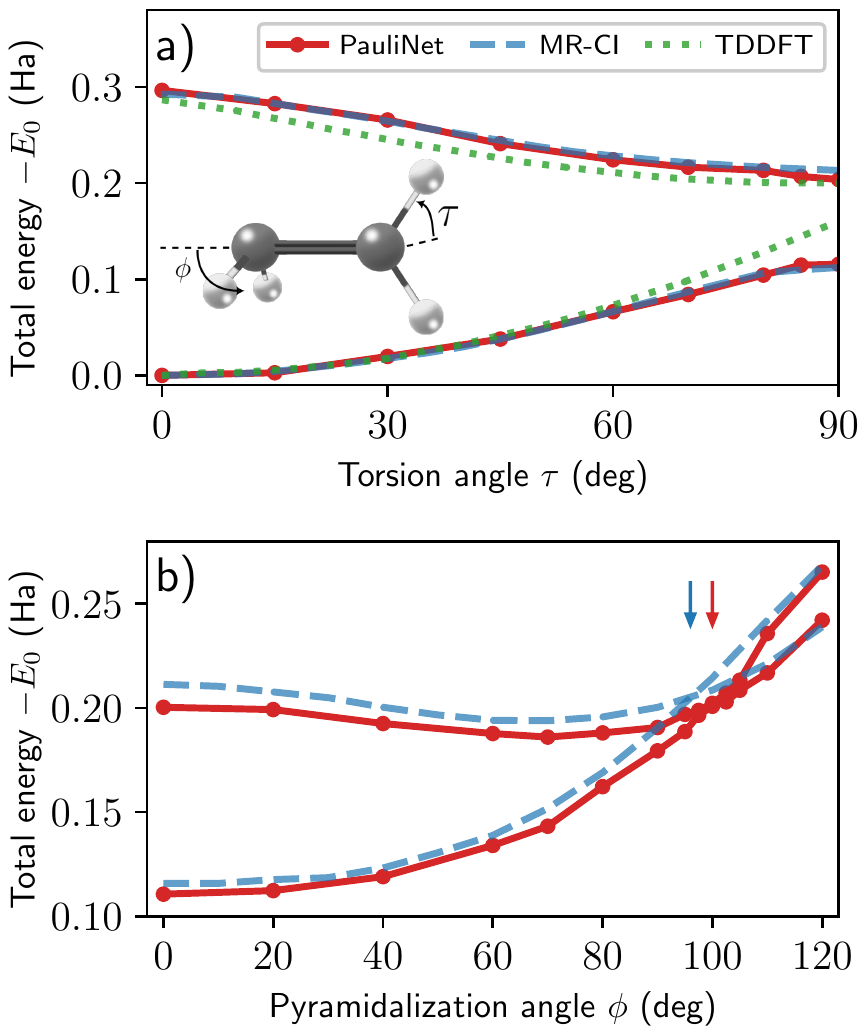}
    \caption{\textbf{Modeling a conical intersection of ethylene.} Inset: Ethylene structure. \textbf{a} Total energies (relative to the ground state of the planar geometry $E_{0}$) of the ground state and first-excited singlet state  as a function of torsion angle $\tau$, with MR-CI \cite{Barbatti2004} and TDDFT \cite{Malis2020} results also plotted for comparison. TDDFT overestimates the barrier (the ground state at $\tau=90 \degree$) and produces an unphysical cusp, while the MR-CI results which predict an avoided crossing are well reproduced by PauliNet. \textbf{b} Same as above but as a function of pyramidalization angle $\phi$ ($\tau=90 \degree$), with the degeneracy of the two states producing a conical intersection. The arrows denote the conical intersection, with PauliNet ($\phi \sim 100 \degree$) closely matching the MR-CI result ($\phi \sim 96 \degree$). Note: The geometric parameters (bond lengths and angles) vary slightly between the torsion and pyramidalization experiments (see ref.~\onlinecite{Barbatti2004}). (The numerical data can be found in Supplementary Information Table V.)}
    \label{fig:ethylene}
\end{figure}

As a final application of excited PauliNet, we compute ground- and excited-state potential energies for ethylene (\ch{H$_{2}$C=CH$_{2}$}) as a function of its torsion and pyramidalization angles (see inset of Fig.~\ref{fig:ethylene}). Twisting around the \ch{C=C} bond raises the energy of the ground state while lowering that of the first-excited singlet state, giving rise to an avoided crossing at a torsion angle $\tau$ of $90 \degree$. From this twisted structure, the energy gap between the two states is further reduced through the pyramidalization of one of the CH$_{2}$ groups, leading to a conical intersection. These potential energy curves, whose modeling is often too challenging for single-reference methods \cite{Krylov2001, Malis2020, Barbatti2014}, have been characterized using multi-reference configuration interaction (MR-CI) methods \cite{Barbatti2004} which we use for comparison. 

We choose the same ground-state (planar) geometry as ref.~\onlinecite{Barbatti2004} (optimized using a small CAS and the aug-cc-pVDZ basis set; see Supplementary Information Table I) and find the excitation energy between the ground state and first-excited singlet state to be within a few mHa of the MR-CI results. As we vary $\tau$, while keeping all other geometric parameters fixed, we find the energy curves to be well reproduced by PauliNet, with an avoided crossing at $\tau=90 \degree$ (panel a of Fig.~\ref{fig:ethylene}; curves symmetric about $\tau=90 \degree$). Single-reference methods, such as TDDFT (see figure), often overestimate the energy of the ground state at $\tau=90 \degree$ (barrier) and produce an unphysical cusp.

Next, we take the same twisted structure ($\tau=90 \degree$) as ref.~\onlinecite{Barbatti2004} (optimized using a small CAS and the aug-cc-pVDZ basis set; see Supplementary Information Table I) and vary the pyramidalization angle $\phi$, while keeping all other geometric parameters fixed. While there is a small discrepancy between PauliNet and the MR-CI results (panel b of Fig.~\ref{fig:ethylene}), the trend of the energy curves is well described, including the correct minimum of the excited-state curve ($\sim 70 \degree$) and the conical intersection (PauliNet: $\phi \sim 100 \degree$; MR-CI: $\phi \sim 96 \degree$). We note that many single-reference methods are unable to even qualitatively describe the conical intersection, instead predicting spurious features \cite{Barbatti2014}.

\section{Discussion}
We have introduced an approach to compute highly accurate excited-state solutions of the electronic Schrödinger equation for molecules by using deep neural networks that are trained in an unsupervised manner with variational Monte Carlo. We have employed the PauliNet architecture \cite{Hermann2020} to approximate the ground- and excited-state wavefunctions, however other architectures such as FermiNet \cite{Pfau2020} or second quantization approaches \cite{Choo2020} could also be employed, with suitable modifications. As our approach to find excited states only constrains the excited-state wavefunctions, the ability to compute highly accurate and variational absolute ground-state energies is unchanged. In addition, we demonstrate for a number of small molecules containing up to 42 electrons, that excited PauliNet can reliably find the first excitation energies with an accuracy that is on par with high-level electronic-structure methods, whereas cheaper methods such as TDDFT are less consistent in approximating these energies. The accuracy of the excited-state wavefunctions is underlined by an accurate match of oscillator strengths, which depend on the transition dipole moment, a quantity that is more sensitive to the exact form of the wavefunction than the energy. For benzene (42 electrons), PauliNet already requires significantly less computational time than higher-order methods, and this advantage will only improve for larger molecules. Formally, a single PauliNet is an $\mathcal{O}(N^{4})$ method for $N$ electrons, due to the computational cost of the Hartree-Fock or CASSCF baseline, however, in practice we empirically observe an $\mathcal{O}(N^{3})$ dependency for the system sizes tested, as discussed above. In addition, for excited-state calculations $n$ PauliNet replicas are used which gives rise to $\mathcal{O}(nN^{3})+\mathcal{O}(n^{2}N^{2})$, with the latter term arising from the pairwise overlaps and having a much smaller prefactor than the former.

Notably, almost identical excited PauliNet architectures are used across the systems shown in this paper -- up to minor modifications such as the budget of Slater determinants and the total number of excited states requested, and a deeper network for benzene to adapt for a potentially more complex wavefunction. Whereas a skilled quantum chemist can usually tune and specialize an existing electronic-structure method to give very high-accuracy results for a given molecule, our aim is the exact opposite: to provide a method that, by leveraging machine learning tools, is as automated as possible and will work over a wide range of Hamiltonians provided.

We have demonstrated that we can compute ground- and excited-state potential energy surfaces with the example of ethylene where we model an avoided crossing and conical intersection. Here, where single-reference methods often fail, PauliNet performs well against multi-reference CI results. By combining the present approach with recent and ongoing extensions of PauliNet \cite{Scherbela2021} and FermiNet \cite{Gao2021} that variationally compute entire potential energy surfaces, both highly accurate ground- and excited-state energy surfaces are now accessible with deep VMC methods. Future work will investigate the application of PauliNet to other interesting processes where molecular dynamics interacts with excited states.

One of the limitations of the current approach is that it appears difficult to reliably find all excited states up to a given desired number, especially in cases where several excited states have similar energies. This is a complex problem which depends on the Hartree-Fock/CASSCF initialization, on the total number of states requested, on the learning algorithm, and the expressiveness of the architecture and will be studied in more detail elsewhere. However, the first excited state could be reliably found for all molecules studied here, and apart from one exception also the second excited state. This, in combination with the high numerical accuracy and the favorable computational cost, makes deep VMC a promising method to compute both ground- and excited-state properties for small- and medium-sized molecules with dozens or even low hundreds of electrons.
\clearpage
\section*{Methods}
\subsection*{PauliNet ansatz}
At the heart of our approach is the PauliNet ansatz, introduced in ref.~\onlinecite{Hermann2020} and further refined in ref.~\onlinecite{Schaetzle2021}, a multi-determinant Slater-Jastrow-backflow type trial wavefunction that is parametrized by highly expressive deep neural networks:
\begin{gather}
\label{eq:pn_ansatz}
\psi_{\boldsymbol\theta}(\boldsymbol{\mathrm r})
  = \mathrm e^{\gamma(\boldsymbol{\mathrm r})+J_{\boldsymbol\theta}(\boldsymbol{\mathrm r})}
  \textstyle \sum\limits_{p} c_p
  \det[\tilde\varphi_{\boldsymbol\theta,{\mu_p}i}^\uparrow(\boldsymbol{\mathrm r})]
  \det[\tilde\varphi_{\boldsymbol\theta,{\mu_p}i}^\downarrow(\boldsymbol{\mathrm r})], \\
  \label{eq:pn_ansatz_2}
  \tilde\varphi_{\boldsymbol\theta,\mu i}(\mathbf r) =\varphi_\mu(\mathbf r_i)f_{\boldsymbol\theta,\mu i}^{(\mathrm m)}(\mathbf r) + f_{\boldsymbol\theta,\mu i}^{(\mathrm a)}(\mathbf r),
\end{gather}
where $\mathbf r = (\mathbf r_{1}, ..., \mathbf r_{N})$ is the $3N$-dimensional real space of electron coordinates. The structure of our ansatz ensures that the correct physics is encoded: the wavefunction obeys exact asymptotic behavior through the fixed electronic cusps $\gamma$, and is antisymmetric with respect to the exchange of like-spin electrons through the use of generalized Slater determinants, guaranteeing the Pauli exclusion principle is obeyed. 

The expressiveness of PauliNet is contained in the Jastrow factor $J_{\boldsymbol\theta}$ and backflow \textbf{f$_{\boldsymbol\theta}$}, which introduce many-body correlation, and are both represented through deep neural networks (denoted by trainable parameters $\boldsymbol\theta$). $J_{\boldsymbol\theta}$ and \textbf{f$_{\boldsymbol\theta}$} are constructed in ways that preserve the antisymmetry of the fermionic wavefunction with respect to exchanging like-spin electrons, as well as its cusp behavior. The Jastrow factor is an exchange-symmetric function, and captures complex correlation effects through augmenting the Slater-determinant baseline, but is incapable of modifying the nodal surface of the determinant expansion. Changes to the nodal surface are possible through the backflow, which acts on the single-electron orbitals $\varphi_\mu$ directly, transforming them into permutation-equivariant many-electron orbitals $\tilde\varphi_\mu$. \textbf{f$_{\boldsymbol\theta}$} is split into multiplicative (m) and additive (a) components (Eq.~\eqref{eq:pn_ansatz_2}), and is designed to be equivariant under the exchange of like-spin electrons.

\subsection*{Ground-state optimization}
Like traditional VMC methods, PauliNet is based on the variational principle, which guarantees that the energy expectation value of a trial wavefunction $\psi_{\boldsymbol\theta}$ is an upper bound to the true ground-state energy: 
\begin{equation}
E_{0} = \underset{\psi}{\text{min}} \langle \psi|\hat{H}|\psi \rangle \leq \underset{\boldsymbol\theta}{\text{min}} \langle \psi_{\boldsymbol\theta}|\hat{H}|\psi_{\boldsymbol\theta} \rangle.
\end{equation}
For a given system, a standard quantum chemistry method (Hartree-Fock (HF) for a single determinant; complete active space self-consistent field (CASSCF) for multiple determinants) is performed, with the solution supplemented by the analytically-known cusp conditions, thus producing a reasonable baseline wavefunction. We then optimize the PauliNet ansatz by minimizing the total electronic energy (serving directly as the loss), following the standard VMC trick of evaluating it as an expectation value of the local energy, $E_{\mathrm{loc}}(\mathbf r) = \hat{H}\psi(\mathbf r)/\psi(\mathbf r)$, over the probability distribution $|\psi_{\boldsymbol\theta}|^{2}$:
\begin{equation}
\label{eq:gs_loss}
    \mathcal{L}(\boldsymbol\theta) = \mathbb{E}_{\mathbf r \sim |\psi_{\boldsymbol\theta}|^{2}}\big[E_{\mathrm{loc}}[\psi_{\boldsymbol\theta}](\mathbf r)\big].
\end{equation}
This means that, in practice, we alternate between sampling electron positions generated using a Langevin algorithm with the probability of the trial wavefunction serving as the target distribution, and optimizing the trial wavefunction parameters using stochastic gradient descent. For further details, see ref.~\onlinecite{Hermann2020}.

\subsection*{Computing excited states}\label{sec:theory_exc}
We now introduce the central idea of this paper: a deep VMC method to compute the ground and low-lying excited states of a given electronic system. While we employ PauliNet to represent the individual wavefunctions, the method can also employ FermiNet or other real-space wavefunction representations with suitable modifications.

In a similar spirit to the ground-state optimization process, we first obtain a reasonable baseline for each state by performing a minimal state-averaged CASSCF calculation. This optimizes the energy average for all states in question and yields a single set of orbitals to construct each multi-determinant wavefunction, which in turn are supplemented by the analytically-known cusp conditions. We fix the number of determinants in our ansatz by cutting off the CASSCF expansion based on the absolute values of their determinant coefficients. The choice of the CASSCF baseline ensures that the PauliNet ansatzes for the different excited states are close to orthogonal upon initialization. In contrast to the ground-state calculation, the optimization of excited states requires a more nuanced choice of the active space. In principle, we must ensure that the solutions contain determinants with orbitals of the necessary rotational symmetries (the Jastrow factor and backflow correction are rotationally-symmetric modifications of the orbitals) and spin configurations (the choice of the number of spin-up and spin-down electrons does impose restrictions on the states that may be attained by our ansatz). For most systems studied in this paper, a generic choice of the active space was sufficient (see Supplementary Table VIII) and we have not studied the dependence on the CAS initialization in more depth. As shown in previous studies the quality of the orbitals has only a minor effect on the training and does not change the final energy \cite{Schaetzle2021}. If, however, the initialization is not accounted for and the baseline solutions provide a qualitatively wrong spectrum of excited states our ansatzes may be trapped in local minima and miss intermediate excited states (see Fig.~\ref{fig:small-systems}), even though we keep the Slater-determinant coefficients $c_p$ and linear coefficients $c_{\mu k}$ of the single-electron orbitals $\varphi_\mu(\mathbf r_i) = \sum_{k} c_{\mu k} \phi_{k}(\mathbf r_i)$ trainable. 

Our objective is to calculate the lowest $n$ eigenstates of a given system, that is, find the set of orthogonal states that minimizes the energy expectation value. We approach this challenge by introducing a penalty term to the energy loss function (Eq.~\eqref{eq:gs_loss}) and optimizing the joint loss for $n$ PauliNet instances:
\begin{equation}\label{eq:loss}
    \mathcal{L}(\boldsymbol\theta) = \underbrace{\sum_{i} \mathbb{E}_{i} \big[E_{\mathrm{loc}}[\psi_{\boldsymbol\theta, i}](\boldsymbol{\mathrm r})\big]}_\text{energy minimization} + \ \alpha \underbrace{\sum_{i>j} \bigg(\frac{1}{1-|S_{ij}|}-1\bigg)}_\text{overlap penalty},
\end{equation}
where $\mathbb{E}_{i} = \mathbb{E}_{\boldsymbol{\mathrm r} \sim |\psi_{\boldsymbol\theta, i}|^2}$ and $S_{ij}$ is the pairwise overlap between states $i$ and $j$. The functional form of the overlap penalty is chosen to diverge when two states collapse and behave linearly when states are close to orthogonal (see the next section for details). This allows states to overlap during the optimization procedure, while preventing their collapse and eventually driving them to orthogonality when they have settled in a local minimum of the energy. The hyperparameter $\alpha$ weights the two loss terms and can be increased throughout the training to strengthen the orthogonality condition when approaching the final wavefunctions. For a sufficiently large $\alpha$ the true minimum of the loss function corresponds to the sum of the energies of the lowest-lying excited states with these states having no overlap. Thus, optimizing the penalized loss function (Eq.~\eqref{eq:loss}) leads to an unbiased convergence towards the lowest-lying excited states (see below). In practice a small $\alpha$ is typically sufficient, making a robust choice possible. 

To stabilize the training and reduce the computational cost we detach gradients in such a way that we only consider the overlap with the lower-lying states respectively, that is, the ground state is subject to unconstrained energy minimization and the $n$-th excited state introduces $n$ pairwise penalty terms. We compute the overlap of the unnormalized states $i$ and $j$ as the geometric mean of the two Monte Carlo estimates, obtained over distributions $|\psi_{\boldsymbol\theta, i}|^2$ and $|\psi_{\boldsymbol\theta, j}|^2$, respectively:

\begin{equation}\label{eq:overlap}
S_{ij} = \mathrm{sgn}\bigg(\mathbb{E}_{i}\bigg[\frac{\psi_{\boldsymbol\theta, j}(\mathbf r)}{\psi_{\boldsymbol\theta, i}(\mathbf r)}\bigg]\bigg) \times\sqrt{\mathbb{E}_{i}\bigg[\frac{\psi_{\boldsymbol\theta, j}(\mathbf r)}{\psi_{\boldsymbol\theta, i}(\mathbf r)}\bigg] \ \mathbb{E}_{j}\bigg[\frac{\psi_{\boldsymbol\theta, i}(\mathbf r)}{\psi_{\boldsymbol\theta, j}(\mathbf r)}\bigg]}.
\end{equation}
The sign of the overlap can be obtained from either of the two estimators, which match in the limit of infinite sampling. If the overlap is close to zero and the signs of the two estimates differ due to statistical noise of the sampling, we consider the states to be orthogonal. Similar to the energy loss, the gradient \footnote{We employ gradient clipping to stabilize the training.} of the pairwise overlap can be formulated such that it depends on the first derivative of the log wavefunction with respect to the parameters only (see below for details).

Finally, we note that different states may be modeled at different levels of quality, which can lead to erroneous excitation energies. In order to improve the error cancellation of our ansatzes we employ a variance-matching technique. As the variance of the energy $\sigma^{2}$ can be considered a metric of how close a wavefunction is to a true eigenstate, variance-matching procedures can be useful tools \cite{PinedaFlores2019, Otis2020, Robinson2017}. Here, we utilize a simple scheme: for single-state quantities such as total energies, we evaluate all wavefunctions at the end of training. For multi-state quantities, such as excitation energies or transition dipole moments, we match states of a similar variance. That is, if final $\psi_{\boldsymbol\theta, i}$ has a lower variance than final $\psi_{\boldsymbol\theta, j}$, we take $\psi_{\boldsymbol\theta, i}$ at an earlier point in training. This simply involves computing $\sigma^{2}$ of the training energies and applying exponential moving average at each iteration to monitor convergence (see below for details). We find this procedure typically improves the final results.
\subsection*{Loss function and overlap penalty}\label{sec:penalty}
\begin{figure}
    \centering
    \includegraphics[width=\columnwidth]{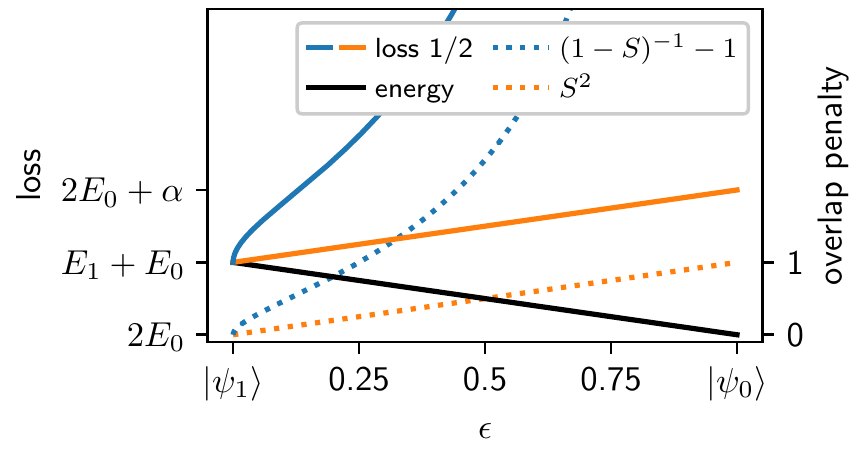}
    \caption{\textbf{Sketch of the loss function.} This figure illustrates the behavior of our loss function for a two-state system. The ground state is kept fixed and the second state is considered to be a linear combination of the ground state and first excited state (Eq.~\eqref{psi_eps}). The scales are to be understood in arbitrary units, as they depend on the choice of the hyperparameters and the energies of the system under investigation.}
    \label{fig:loss_penalty}
\end{figure}

There are a number of choices of possible loss functions for the optimization of excited states in quantum Monte Carlo \cite{Pathak2021,Garner2020,Dash2019}. In order to assess the feasibility of excited-state optimization with deep neural network ansatzes in variational Monte Carlo we conducted a range of experiments with different types of optimization objectives. Our empirical findings showed that employing a penalty method is the conceptually most straightforward approach and gives stable results when combining it with our implementation of PauliNet. Initially, we started with an overlap penalty term similar to Pathak et al\cite{Pathak2021}. However, we found that our optimization could still collapse even if we chose a sufficiently large prefactor ($\alpha$) and the training could not recover. We therefore switched to an alternative penalty term (Eq.~\eqref{eq:loss}) which diverges upon a collapse of the states. The effect of our penalty term can be illustrated by considering the loss for a two-state system with the exact ground state $|\psi_0 \rangle$ and a linear combination of the ground and first excited state $|\psi_1 \rangle$ (see Fig.~\ref{fig:loss_penalty}):
\begin{equation}\label{psi_eps}
    |\psi_\epsilon \rangle = \sqrt{1-\epsilon} | \psi_1 \rangle + \sqrt{\epsilon} | \psi_0 \rangle.
\end{equation}
The overlap and the energy can be obtained as 
\begin{align}
    \langle \psi_0|\psi_\epsilon\rangle =\sqrt{\epsilon}, ~~~ \langle \psi_\epsilon|H|\psi_\epsilon\rangle = (1-\epsilon)E_1 + \epsilon E_0.
\end{align}
In the vicinity of the orthogonal solution, the Taylor expansion of the penalty term is 
\begin{equation}
    \frac{1}{1-|S|}-1 = |S| + |S|^2 + |S|^3 + ..., ~ \text{at} ~~|S| = 0,
\end{equation}
that is, the overlap penalty behaves linearly to first order. This gives rise to a penalty that is locally stable for any prefactor, lower bounded by the $S^2$ penalty term, and diverges if states collapse. For a large enough $\alpha$ parameter the global optimum of the total loss is at zero overlap, that is, the optimization method is incentivized to find exactly orthogonal states without mixing. 

\begin{figure}
    \centering
    \includegraphics[width=\columnwidth]{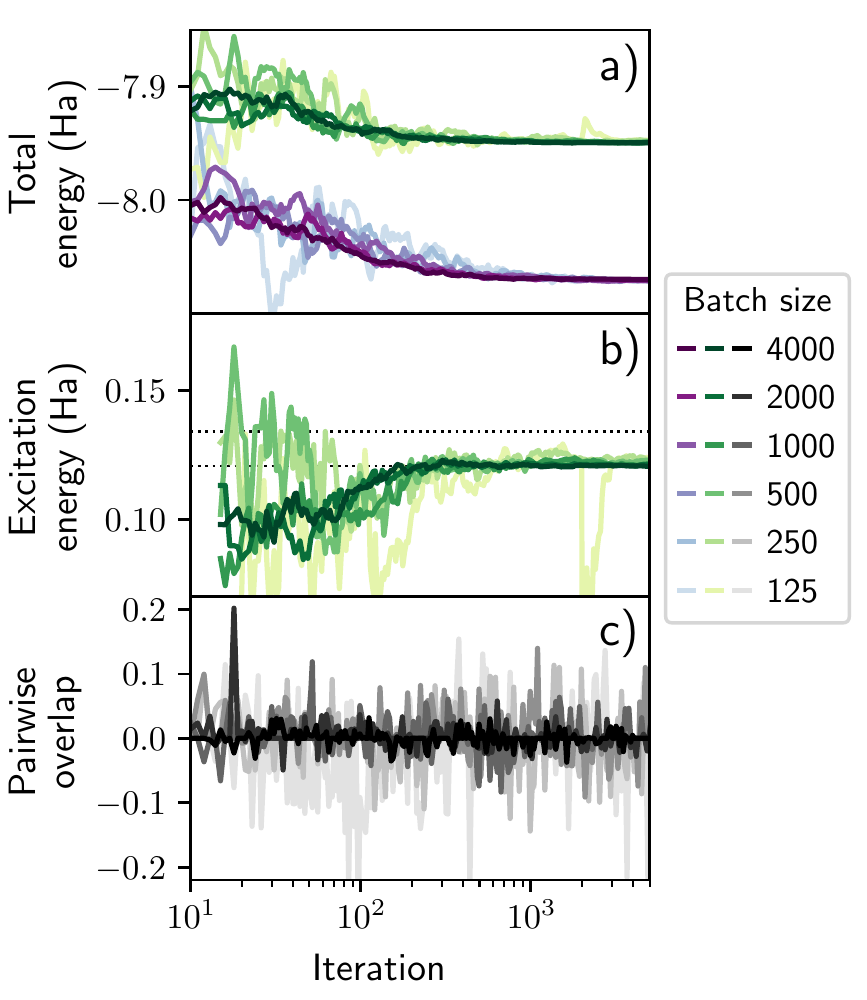}
    \caption{\textbf{Behavior of the loss function with batch size.} The ground state and first excited state of LiH are approximated. The convergence of the total energies \textbf{(a)}, excitation energy \textbf{(b)} and the pairwise overlap between the wavefunctions \textbf{(c)} are shown. Dotted horizontal lines are excitation energies from highly accurate references \cite{bandeLiHPotentialEnergy2010, jasikElectronicStructureTimedependent2017}. While the optimization works well for the large batches that we typically employ in our calculations ($\geq 2000$), this becomes less reliable, at least for the excited state, in the limit of smaller batch sizes. Note: Darker corresponds to a larger batch size in each respective color.}
    \label{fig:loss_batch}
\end{figure}

In practice, for the batch sizes used in our calculations, we have not observed a bias due to the non-linear nature of the penalty when applied to sampled expectation values of the overlap. However, it is expected that this is no longer the case in the limit of small batches. In order to elucidate how our loss function behaves in this regard, we compute the two lowest states of LiH using a range of different batch sizes (see Fig.~\ref{fig:loss_batch}). We find the optimization procedure to be robust for the large batch sizes that we typically employ ($\geq 2000$), with the excitation energy within 1~mHa of the exact, and the pairwise overlap remaining small throughout training (panel c). For smaller batch sizes, we observe a larger degree of statistical noise in the pairwise overlap, which leads to a less reliable approximation for the excited state and the corresponding excitation energy (panel b).

\subsection*{Gradient of the loss function}
In order to differentiate the loss function we explicitly formulate the gradient. We consider the general case of a mixed observable:
\begin{align}
    O_{ij} & = \frac{1}{N_{i}N_{j}} \int d^{3}\mathbf r \ \psi_{\boldsymbol\theta, i}(\mathbf r) \big[\hat{O} \psi_{\boldsymbol\theta, j}(\mathbf r)\big], \\
    & = \frac{N_{i}}{N_{j}} \mathbb{E}_{i}\bigg[\frac{\hat{O}\psi_{\boldsymbol\theta, j}(\mathbf r)}{\psi_{\boldsymbol\theta, i}(\mathbf r)}\bigg],
\end{align}
where $N_{i}$, $N_{j}$ are the norms of the wavefunctions and $\mathbb{E}_{i} = \mathbb{E}_{\boldsymbol{\mathrm r} \sim |\psi_{\boldsymbol\theta, i}|^2}$. By the property of Hermitian matrices, $O_{ij}$ = $O_{ji}$, we derive an expression that does not depend on the wavefunction norms:
\begin{align}
        O_{ij} &= \sqrt{\frac{N_{i}}{N_{j}} \mathbb{E}_{i}\bigg[\frac{\hat{O}\psi_{\boldsymbol\theta, j}(\mathbf r)}{\psi_{\boldsymbol\theta, i}(\mathbf r)}\bigg]} \sqrt{\frac{N_{j}}{N_{i}} \mathbb{E}_{j}\bigg[\frac{\hat{O}\psi_{\boldsymbol\theta, i}(\mathbf r)}{\psi_{\boldsymbol\theta, j}(\mathbf r)}\bigg]}, \nonumber \\ \\
        &=\mathrm{sgn} \label{eq:mixed_operator} \bigg(\mathbb{E}_{i}\bigg[\frac{\hat{O}\psi_{\boldsymbol\theta, j}(\mathbf r)}{\psi_{\boldsymbol\theta, i}(\mathbf r)}\bigg]\bigg) \nonumber \\ 
        & \ \ \ \ \ \ \ \ \times\sqrt{\mathbb{E}_{i}\bigg[\frac{\hat{O}\psi_{\boldsymbol\theta, j}(\mathbf r)}{\psi_{\boldsymbol\theta, i}(\mathbf r)}\bigg] \ \mathbb{E}_{j}\bigg[\frac{\hat{O}\psi_{\boldsymbol\theta, i}(\mathbf r)}{\psi_{\boldsymbol\theta, j}(\mathbf r)}\bigg]}.
\end{align}

This expression reduces to the pairwise overlaps (Eq.~\eqref{eq:overlap}) upon setting $\hat{O}=\text{Id}$. The derivative of this term can be expressed as
\begin{align}
    \partial O_{ij} = &\frac{1}{O_{ij}} \bigg \{\mathbb{E}_{i} \bigg[\bigg(\frac{\hat{O}\psi_{\boldsymbol\theta, j}(\mathbf r)}{\psi_{\boldsymbol\theta, i}(\mathbf r)}-\mathbb{E}_{i}\bigg[\frac{\hat{O}\psi_{\boldsymbol\theta, j}(\mathbf r)}{\psi_{\boldsymbol\theta, i}(\mathbf r)}\bigg]\bigg)\nonumber \\&\partial \ln|\psi_{\boldsymbol\theta, i}(\mathbf r)|\bigg] \times \mathbb{E}_{j}\bigg[\frac{\hat{O}\psi_{\boldsymbol\theta, i}(\mathbf r)}{\psi_{\boldsymbol\theta, j}(\mathbf r)}\bigg] + (i \Longleftrightarrow j)\bigg\},
\end{align}
where $(i \Longleftrightarrow j)$ is an additional term with the two indices interchanged.

By considering the Hamiltonian operator $\hat{H}$ and setting $i = j$ we recover the gradient of the energy loss \cite{Hermann2020}:
\begin{equation}
    \partial E_{ii} = 2 \mathbb{E}_{i}\bigg[\bigg(\frac{\hat{H}\psi_{\boldsymbol\theta, i}(\mathbf r)}{\psi_{\boldsymbol\theta, i}(\mathbf r)} - \mathbb{E}_{i}\bigg[\frac{\hat{H}\psi_{\boldsymbol\theta, i}(\mathbf r)}{\psi_{\boldsymbol\theta, i}(\mathbf r)}\bigg]\bigg)\partial \ln |\psi_{\boldsymbol\theta, i}(\mathbf r)|\bigg].
\end{equation}

\subsection*{Variance matching}\label{sec:variance_matching}
\begin{figure}[H]
    \centering
    \includegraphics[width=\columnwidth]{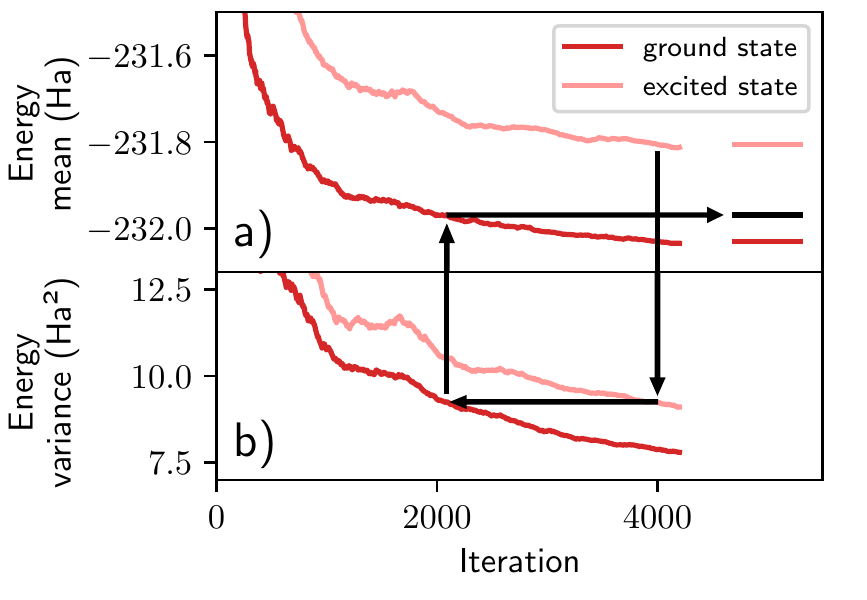}
    \caption{\textbf{Sketch of the variance-matching procedure.} The excitation energy of the benzene calculation at step 4000 is obtained for illustration purposes. The variance (b) of the excited state is higher than that of the ground state and is therefore matched with the variance of the ground state at a previous iteration. The excitation energy is computed by comparing the mean energies (a) at the respective iterations. This acts to reduce the excitation energy and is found to improve the results in all of our experiments.}
    \label{fig:variance_matching}
\end{figure}

As far as relative energies are concerned most computational chemistry methods rely heavily on the cancellation of error. While quantum Monte Carlo methods using neural network-based trial wavefunctions provide highly accurate total energies, the flexibility of these ansatzes is difficult to control which can lead to varying qualities of approximations for different states. In order to account for potential imbalances we utilize the variance of the wavefunctions as a measure of the quality of the approximation (zero-variance principle) and employ a variance-matching scheme. Variance-matching techniques as well as variance extrapolation have typically been applied by optimizing a family of ansatzes and comparing variances across the optimized wavefunctions \cite{Robinson2017}. Instead of training multiple ansatzes we checkpoint wavefunctions during the training and compute excitation energies by rewinding the ground state to match the variance of the excited state as depicted in Fig.~\ref{fig:variance_matching}. The mean and variance of each wavefunction are computed over the batch dimension at each step in training and smoothed with an exponential walking average. For the final estimation of excitation energies, the respective wavefunctions are then sampled exhaustively as in the usual evaluation process. While the variance matching hardly impacts the excitation energies for small systems, for larger and harder to optimize systems, such as benzene, it becomes increasingly relevant.

\subsection*{Spin treatment} \label{sec:spin_treatment}
PauliNet encodes only the spatial part of the wavefunction and its like-spin antisymmetry explicitly \cite{Foulkes2001}, while the spin part, which guarantees the opposite-spin antisymmetry, is only implicit. Every spin-assigned spatial ansatz such as PauliNet is always an eigenstate of $\mathcal S_z$ with an eigenvalue of $M=\frac12(N_\uparrow-N_\downarrow)$, but it may not be an eigenstate of $\mathcal S^2$. The spatial part of eigenstates of $\mathcal S^2$ is characterized by specific sets of permutational symmetries involving opposite-spin electrons \cite{Pauncz79}. PauliNet does not enforce these symmetries but instead attempts to learn them through the variational principle because eigenstates of the Hamiltonian are also eigenstates of $\mathcal S^2$. Therefore, we do not, in general, control the spin of the eigenstates found in the optimization procedure --- they are simply found in the order of increasing energy, independent of spin. The spin of a found eigenstate can be obtained in principle by Monte Carlo sampling \cite{HuangJCP98}. Whether a particular spin state is found in practice may be influenced by the spin of the CASSCF baseline wavefunction, which we therefore report in Supplementary Table VIII. In special cases, we may wish to target a specific spin state (e.g., see the section \enquote{Multi-reference application: conical intersections}), and for that, we can take advantage of the orbital-assigned backflow of PauliNet. Combined with the freezing of the determinant coefficients, this ensures that PauliNet remains in the same spin state as the CASSCF baseline wavefunction.

\section*{Data Availability}
\noindent The dataset generated in this study is openly available in Zenodo (https://doi.org/10.5281/zenodo.7274855). Source data are provided with this paper.

\section*{Code Availability}
\noindent The computer code used in this study is openly available in Zenodo (https://doi.org/10.5281/zenodo.7347937).

\clearpage

\bibliographystyle{apsrev4-2}
\bibliography{mainReferences}

\section*{Acknowledgments}
\noindent We thank Tim Gould (Griffith U) for early discussions about variational principles for excited states.
Funding is gratefully acknowledged from the Berlin mathematics center MATH+ (Projects AA1-6, AA2-8),  European Commission (ERC CoG 772230), Deutsche Forschungsgemeinschaft (NO825/3-2), and the Berlin Institute for Foundations in Learning and Data (BIFOLD).

\section*{Author contributions}
\noindent M.T.E, Z.S, J.H and F.N designed the research. M.T.E, Z.S, P.A.E, J. H and F.N developed the method. M.T.E, Z.S and J.H wrote the computer code. M.T.E and Z.S carried out the numerical calculations. M.T.E, Z.S, P.A.E, J.H and F.N analyzed the data. M.T. E, Z.S, J.H and F.N wrote the manuscript.

\section*{Competing interests}
\noindent The authors declare no competing interests.

\section*{Additional information}
\noindent \textbf{Supplementary material} is available below. \\ \\
\noindent \textbf{Correspondence} and requests for materials should be addressed to J. Hermann or F. Noé.

\newpage
\onecolumngrid
\appendix
\renewcommand{\tablename}{Supplementary Table}
\section{Geometries of test systems}
All calculations presented in Sections II A -- C of the main paper were done at the ground-state equilibrium geometries. The geometries for all molecules are listed in Supplementary Table~\ref{table:geometries}. For the ethylene structures presented in Section II D, we list the planar (ground-state equilibrium) and twisted ($\tau=90$) geometries.

\begin{table}[H]
\caption{Ground-state equilibrium geometries of the test systems.}\label{table:geometries}
\begin{tabular*}{\columnwidth}{@{\extracolsep\fill}lcclcc}
\hline\hline
  \T Molecule & Atom & Position (\AA) & Molecule & Atom & Position (\AA)\B\\ 
\hline
  LiH  & Li & (0.000, 0.000, 0.000) & BeH & Be & (0.000000, 0.000000, 0.000000)\T\\ 
       & H & (1.595, 0.000, 0.000) & & H & (1.326903, 0.000000, 0.000000)\B\\ 
\hline
  BH   & B & (0.000000, 0.000000, 0.000000) & CH$^{+}$ & C & (0.00000, 0.00000, 0.00000)\T\\ 
       & H & (0.000000, 0.000000, 1.222874) & & H & (1.13092, 0.00000, 0.00000)\B\\
\hline
  H$_{2}$O   & O & (0.000000, 0.000000, -0.069903) & CO & C & (0.000000, 0.000000, -0.661165)\T\\ 
             & H & (0.000000, 0.757532, 0.518435) & & O & (0.000000, 0.000000, 0.472379)\\
             & H & (0.000000, -0.757532, 0.518435)\B\\
\hline
  NH$_{3}$   & N & (0.067759, -0.000000, 0.000000) & C$_{2}$H$_{4}$ (planar) & C & (-0.675000, 0.000000, 0.000000)\T\\ 
             & H & (-0.313823, 0.468746, -0.811891) & & C & (0.675000, 0.000000, 0.000000)\\ 
             & H & (-0.313823, -0.937491, -0.000000) & & H & (-1.242900, 0.000000, -0.930370) \\ 
             & H & (-0.313823, 0.468746, 0.811891) & & H & (-1.242900, 0.000000, 0.930370)\\  
             & & & & H & (1.242900, 0.000000, -0.930370)\\    
             & & & & H & (1.242900, 0.000000, 0.930370)\B\\                 
\hline
  C$_{2}$H$_{4}$ (twisted) & C & (-0.688500, 0.000000, 0.000000) & C$_{6}$H$_{6}$ & C & (0.000000, 1.396792, 0.000000)\T\\ 
  & C & (0.688500, 0.000000, 0.000000) & & C & (0.000000, -1.396792, 0.000000) \\
  & H & (-1.307207, 0.000000, -0.915547) & & C & (1.209657, 0.698396, 0.000000) \\ 
  & H & (-1.307207, 0.000000, 0.915547) & & C & (-1.209657, -0.698396, 0.000000) \\ 
  & H & (1.307207, -0.915547, 0.000000) & & C & (-1.209657, 0.698396, 0.000000) \\ 
  & H & (1.307207, 0.915547, 0.000000) & & C & (1.209657, -0.698396, 0.000000) \\ 
  & & & & H & (0.000000, 2.484212, 0.000000) \\ 
  & & & & H & (2.151390, 1.242106, 0.000000) \\ 
  & & & & H & (-2.151390, -1.242106, 0.000000) \\ 
  & & & & H & (-2.151390, 1.242106, 0.000000) \\ 
  & & & & H & (2.151390, -1.242106, 0.000000) \\ 
  & & & & H & (0.000000, -2.484212, 0.000000)\B\\
\hline\hline
\end{tabular*}
\end{table}

\newpage

\section{Energies for small atoms and molecules}
We tabulate the total energies and vertical excitation energies that were obtained for the small systems, as presented in Section II A of the main paper. These are listed in Supplementary Table~\ref{table:small_atoms}.

\begin{table*}[htbp]
\caption{Total energies and vertical excitation energies (both in Ha) for the small atoms in Section II A.}\label{table:small_atoms}
\begin{ruledtabular}
\begin{tabular}{@{\extracolsep\fill}lcccc}
 \T System  & $E_{n}$ & Correlation energy (\%) & $\Delta E$ & $\Delta E \ (\sigma^{2} \ \text{matching})$\B\\
  \hline
  Li & -7.47800(4) & 99.9(1)\footnotemark[1]  & --- & ---\T\\
  & -7.40998(5) & --- & 0.06801(7) & 0.06807(7)\\
  & -7.40988(7) & --- & 0.06812(8) & 0.06811(9)\\
  & -7.40977(7) & --- & 0.06823(8) & 0.06824(9)\\
  & -7.335(4) & --- & 0.143(4) & 0.143(4)\B\\
  \hline
  Be & -14.66736(7) & 100.0(1)\footnotemark[1]  & --- & ---\T\\
  & -14.5663(6) & --- & 0.1011(6) & 0.1009(6)\\
  & -14.5664(1) & --- & 0.1010(1) & 0.1007(2)\\
  & -14.5664(1) & --- & 0.1010(1) & 0.1004(2)\\
  & -14.4710(2) & --- & 0.1964(2) & 0.1952(2)\B\\
  \hline
  B & -24.6509(2) & 97.6(1)\footnotemark[1]  & --- & ---\T\\
  & -24.6506(2) & --- & 0.0004(3) & 0.0005(3)\\
  & -24.6498(2) & --- & 0.0011(3) & 0.0008(3)\\
  & -24.5206(2) & --- & 0.1303(2) & 0.1300(3)\\
  & -24.5202(2) & --- & 0.1307(3) & 0.1300(3)\B\\
  \hline
  C & -37.8388(3) & 96.0(2)\footnotemark[1]  & --- & ---\T\\
  & -37.8388(2) & --- & -0.0001(4) & 0.0001(4)\\
  & -37.8378(2) & --- & 0.0010(4) & 0.0008(4)\\
  & -37.7918(2) & --- & 0.0470(4) & 0.0470(4)\\
  & -37.7911(3) & --- & 0.0477(4) & 0.0451(4)\\
  \hline    
  N & -54.5836(3) & 97.0(2)\footnotemark[1]  & --- & ---\T\\
  & -54.4911(4) & --- & 0.0925(5) & 0.0900(5)\\
  & -54.4885(4) & --- & 0.0952(5) & 0.0893(6)\\
  & -54.4874(4) & --- & 0.0963(6) & 0.0883(6)\\
  & -54.4861(4) & --- & 0.0975(5) & 0.0918(6)\\
  \hline
  O & -75.0532(5) & 94.6(2)\footnotemark[1]  & --- & ---\T\\
  & -75.0506(5) & --- & 0.0027(7) & -0.0003(8)\\
  & -75.0519(5) & --- & 0.0013(7) & -0.0011(7)\\
  & -74.9763(6) & --- & 0.0769(8) & 0.0711(8)\\
  & -74.9771(6) & --- & 0.0761(8) & 0.0703(9)\B\\
\end{tabular}
\end{ruledtabular}
\begin{minipage}{0.9\linewidth}
\footnotetext[1]{Exact energy and HF energy at the CBS limit \cite{Darragh2005}}
\end{minipage}
\end{table*}

\begin{table*}[htbp]
\renewcommand\thetable{II (continued)}
\caption{Total energies and vertical excitation energies (both in Ha) for the many-state calculations in Section II A.}\label{table:small_atoms_continued}
\begin{ruledtabular}
\begin{tabular}{@{\extracolsep\fill}lcccc}
 \T System  & $E_{n}$ & Correlation energy (\%) & $\Delta E$ & $\Delta E \ (\sigma^{2} \ \text{matching})$\B\\
  \hline
  LiH & -8.0695(1) & 98.9(1)\footnotemark[1]  & --- & ---\T\\
  & -7.9497(1) & --- & 0.1198(1) & 0.1199(1)\\
  & -7.9348(1) & --- & 0.1348(2) & 0.1347(2)\\
  & -7.9112(1) & --- & 0.1584(2) & 0.1573(2)\\
  & -7.9114(1) & --- & 0.1582(1) & 0.1574(2)\\
  & -7.8978(1) & --- & 0.1717(2) & 0.1710(2)\\
  & -7.8977(1) & --- & 0.1718(2) & 0.1709(2)\\
  & -7.8549(1) & --- & 0.2147(2) & 0.2130(2)\B\\
  \hline
  BeH & -15.2452(4) & 98.3(4)\footnotemark[1]  & --- & ---\T\\
  & -15.1516(2) & --- & 0.0936(4) & 0.0916(3)\\
  & -15.1507(2) & --- & 0.0946(5) & 0.0914(3)\\
  & -15.0286(3) & --- & 0.2166(5) & 0.2132(4)\\
  & -15.0259(2) & --- & 0.2193(5) & 0.2156(4)\\
  & -15.0249(2) & --- & 0.2203(5) & 0.2138(17)\\
  & -14.9951(4) & --- & 0.2500(5) & 0.2379(5)\\
  & -14.9528(3) & --- & 0.2923(5) & 0.2793(16)\B\\
 \hline
  Be & -14.6667(1) & 99.3(1)\footnotemark[1]  & --- & ---\T\\
  & -14.5655(2) & --- & 0.1012(2) & 0.1011(2)\\
  & -14.5654(1) & --- & 0.1013(2) & 0.1007(2)\\
  & -14.5651(2) & --- & 0.1016(2) & 0.1009(3)\\
  & -14.4663(3) & --- & 0.2004(4) & 0.1967(10)\\
  & -14.4671(3) & --- & 0.1996(3) & 0.1976(5)\\
  & -14.4665(3) & --- & 0.2002(3) & 0.1982(5)\\
  & -14.3915(2) & --- & 0.2752(2) & 0.2732(4)\B\\
\end{tabular}
\end{ruledtabular}
\begin{minipage}{0.9\linewidth}
\footnotetext[1]{Exact energy and HF energy at the CBS limit \cite{Darragh2005}}
\end{minipage}
\end{table*}
\setcounter{table}{2}

\clearpage

\section{Energies and oscillator strengths for intermediate systems}
We tabulate the total energies, vertical excitation energies and oscillator strengths that were obtained for the intermediate systems, as presented in Section II B of the main paper. These are listed in Supplementary Table~\ref{table:intermediate}.
\begin{table*}[htbp]
\caption{Total energies (in Ha), vertical excitation energies (in Ha) and oscillator strengths (in au) for the systems in Section II B.}\label{table:intermediate}
\begin{ruledtabular}
\begin{tabular}{@{\extracolsep\fill}lcccccc}
 \T Molecule  & $E_{n}$ & Correlation energy (\%) & $\Delta E$ & $\Delta E \ (\sigma^{2} \ \text{matching})$ & $f$ & $f  \ (\sigma^{2} \ \text{matching})$\B\\
  \hline
  BH & -25.2883(2) & 99.5(1)\footnotemark[1]  & --- & --- & --- & ---\T\\
  & -25.2361(2) & --- & 0.0522(3) & 0.0511(3) & --- & ---\\
  & -25.2356(2) & --- & 0.0527(3) & 0.0503(3) & --- & ---\\
  & -25.1765(2) & --- & 0.1118(3) & 0.1088(3) & 0.0276(1) & 0.0253(1)\B\\
  \hline
  CH$^{+}$ & -38.0863(2) & --- & --- & --- & --- & ---\T\\
  & -38.0385(2) & --- & 0.0478(3) & 0.0460(3) & --- & ---\\
  & -38.0375(3) & --- & 0.0488(4) & 0.0465(4) & --- & ---\\
  & -37.9664(3) & --- & 0.1199(4) & 0.1172(4) & 0.00606(4) & 0.00576(3)\B\\
  \hline
  H$_{2}$O & -76.4230(5) & 95.9(1)\footnotemark[2] & --- & --- & --- & ---\T\\
  & -76.1499(6) & --- & 0.2731(8) & 0.2702(8) & --- & ---\\
  & -76.1343(6) & --- & 0.2887(8) & 0.2845(9) & 0.0395(3) & 0.0374(3)\B\\
  \hline   
  NH$_{3}$ & -56.5533(3) & 96.6(1)\footnotemark[2] & --- & --- & --- & ---\T\\
  & -56.3122(4) & --- & 0.2411(5) & 0.2391(6) & --- & ---\\
  & -56.2972(4) & --- & 0.2561(5) & 0.2531(6) & 0.0630(4) & 0.0731(5)\B\\
  \hline  
  CO & -113.3039(6) & 95.9(1)\footnotemark[2] & --- & --- & --- & ---\T\\
  & -113.0618(7) & --- & 0.2421(9) & 0.2386(9) & --- & ---\\
  & -112.9765(7) & --- & 0.3274(9) & 0.3211(9) & 0.1048(5) & 0.1030(5)\B\\
\end{tabular}
\end{ruledtabular}
\begin{minipage}{0.9\linewidth}
\footnotetext[1]{Exact energy \cite{Giner2021} and HF energy at the CBS limit \cite{NIST}}
\footnotetext[2]{Exact energy and HF energy at the CBS limit \cite{Darragh2005}}
\end{minipage}
\end{table*}

\section{Energies for benzene}
We tabulate the total energies and vertical excitation energies that were obtained for benzene, as presented in Section II C of the main paper. These are listed in Supplementary Table~\ref{table:benzene}.

\begin{table*}[htbp]
\caption{Total energies and vertical excitation energies (both in Ha) for benzene in Section II C.}\label{table:benzene}
\begin{ruledtabular}
\begin{tabular}{@{\extracolsep\fill}lccc}
 \T $E_{0}$ & $E_{1}$ & $\Delta E$ & $\Delta E \ (\sigma^{2} \ \text{matching})$\B\\
  \hline
  -232.0675(11) & -231.8628(9) & 0.2047(14) & 0.1637(13)\T\B\\
\end{tabular}
\end{ruledtabular}
\end{table*}

\clearpage

\section{Energies for ethylene}
We tabulate the total energies and vertical excitation energies that were obtained for all ethylene structures, as presented in Section II D of the main paper. These are listed in Supplementary Table~\ref{table:ethylene}.

\begin{table*}[htbp]
\caption{Total energies and vertical excitation energies (both in Ha) for ethylene in Section II D.}\label{table:ethylene}
\begin{ruledtabular}
\begin{tabular}{@{\extracolsep\fill}lccclccc}
 \T $\tau \ (\phi=0)$ & $E_{n}$ & $\Delta E$ & $\Delta E \ (\sigma^{2} \ \text{matching})$ & $\phi \ (\tau=90\degree)$ & $E_{n}$ & $\Delta E$ & $\Delta E \ (\sigma^{2} \ \text{matching})$\B\\
  \hline
  0  & -78.5650(4) & --- & --- & 0 & -78.4477(4) & --- & ---\T\\
     & -78.2617(5) & 0.3033(6) & 0.2964(7) &   & -78.3578(5) & 0.0899(6) & 0.0897(7)\\
  15 & -78.5609(4) & --- & --- & 20 & -78.4471(4) & --- & ---\\
     & -78.2755(5) & 0.2854(6) & 0.2799(7) &   & -78.3589(5) & 0.0882(6) & 0.0869(7)\\    
  30 & -78.5461(4) & --- & --- & 40 & -78.4391(4) & --- & ---\\
     & -78.2925(5) & 0.2536(6) & 0.2459(7) &   & -78.3661(5) & 0.0730(6) & 0.0735(6)\\    
  45 & -78.5254(4) & --- & --- & 60 & -78.4267(5) & --- & ---\\
     & -78.3171(5) & 0.2083(6) & 0.2032(7) &   & -78.3704(5) & 0.0563(7) & 0.0537(7)\\    
  60 & -78.4952(4) & --- & --- & 70 & -78.4161(5) & --- & ---\\
     & -78.3339(5) & 0.1613(6) & 0.1580(7) &   & -78.3721(5) & 0.0440(7) & 0.0428(7)\\    
  70 & -78.4771(4) & --- & --- & 80 & -78.3992(4) & --- & ---\\
     & -78.3416(5) & 0.1355(6) & 0.1324(7) &   & -78.3701(5) & 0.0291(6) & 0.0258(7)\\    
  80 & -78.4540(5) & --- & --- & 90 & -78.3828(4) & --- & ---\\
     & -78.3449(5) & 0.1091(7) & 0.1089(7) &   & -78.3675(5) & 0.0153(6) & 0.0112(7)\\    
  85 & -78.4483(4) & --- & --- & 95 & -78.3697(4) & --- & ---\\
     & -78.3512(5) & 0.0971(6) & 0.0922(7) &   & -78.3612(5) & 0.0085(6) & 0.0083(7)\\    
  90 & -78.4425(5) & --- & --- & 97.5 & -78.3634(5) & --- & ---\\
     & -78.3531(5) & 0.0894(7) & 0.0881(7) &   & -78.3593(5) & 0.0041(7) & 0.0023(7)\\    
  & & & & 100 & -78.3587(5) & --- & ---\\  
  & & & &   & -78.3559(5) & 0.0028(7) & 0.0014(7)\\    
  & & & & 102.5 & -78.3552(5) & --- & ---\\  
  & & & &   & -78.3529(5) & 0.0023(7) & 0.0042(7)\\    
  & & & & 105 & -78.3509(5) & --- & ---\\  
  & & & &   & -78.3448(5) & 0.0061(7) & 0.0048(7)\\    
  & & & & 110 & -78.3449(5) & --- & ---\\  
  & & & &   & -78.3224(5) & 0.0225(7) & 0.0189(7)\\      
  & & & & 120 & -78.3284(5) & --- & ---\\
  & & & &   & -78.2929(5) & 0.0355(7) & 0.0231(7)\B\\    
\end{tabular}
\end{ruledtabular}
\end{table*}
\newpage
\section{Overlaps of trained wavefunctions}
For each system we compute the pairwise overlaps between all trained wavefunctions and tabulate the mean. These are listed in Supplementary Table~\ref{table:overlaps}.

\begin{table*}[htbp]
\caption{Mean pairwise overlaps for all systems.}\label{table:overlaps}
\begin{ruledtabular}
\begin{tabular}{@{\extracolsep\fill}lclclclc}
 \T System & $\overline{|S_{ij}|}$ & System & $\overline{|S_{ij}|}$ & System & $\overline{|S_{ij}|}$ & System & $\overline{|S_{ij}|}$\B\\
  \hline
  Li & 0.0010(6) & CH$^{+}$ & 0.0018(7) & C$_{2}$H$_{4}$ ($\tau=80$) & 0.002(2) & C$_{2}$H$_{4}$ ($\phi=95$) & 0.002(1)\T\\
  Be & 0.004(1) & H$_{2}$O & 0.007(3) & C$_{2}$H$_{4}$ ($\tau=85$) & 0.001(1) & C$_{2}$H$_{4}$ ($\phi=97.5$) & 0.003(2) \\
  B & 0.0023(9) & NH$_{3}$ & 0.014(3) & C$_{2}$H$_{4}$ ($\tau=90$) & 0.008(2) & C$_{2}$H$_{4}$ ($\phi=100$) & 0.002(3) \\  
  C & 0.0030(4) & CO & 0.003(1) & C$_{2}$H$_{4}$ ($\phi=0$) & 0.009(1) & C$_{2}$H$_{4}$ ($\phi=102.5$) & 0.007(1) \\      
  N & 0.0040(5) & C$_{2}$H$_{4}$ ($\tau=0$) & 0.011(3) & C$_{2}$H$_{4}$ ($\phi=20$) & 0.002(2) & C$_{2}$H$_{4}$ ($\phi=105$) & 0.007(2) \\
  O & 0.0035(4) & C$_{2}$H$_{4}$ ($\tau=15$) & 0.007(3) & C$_{2}$H$_{4}$ ($\phi=40$) & 0.003(2) & C$_{2}$H$_{4}$ ($\phi=110$) & 0.009(1) \\
  LiH & 0.0047(9)&  C$_{2}$H$_{4}$ ($\tau=30$) & 0.004(3) & C$_{2}$H$_{4}$ ($\phi=60$) & 0.001(1) & C$_{2}$H$_{4}$ ($\phi=120$) & 0.002(3) \\
  BeH & 0.0068(5) & C$_{2}$H$_{4}$ ($\tau=45$) & 0.003(2) & C$_{2}$H$_{4}$ ($\phi=70$) & 0.002(2) & C$_{6}$H$_{6}$ & 0.001(4) \\
  Be (many states) & 0.0035(5) & C$_{2}$H$_{4}$ ($\tau=60$) & 0.008(2) & C$_{2}$H$_{4}$ ($\phi=80$) & 0.002(3) & & \\
  BH & 0.0033(9) & C$_{2}$H$_{4}$ ($\tau=70$) & 0.002(2) & C$_{2}$H$_{4}$ ($\phi=90$) & 0.006(2) & & \B\\    
\end{tabular}
\end{ruledtabular}
\end{table*}

\clearpage

\section{Hyperparameters}
We tabulate the hyperparameters that were used in all calculations. These are listed in Supplementary Table~\ref{table:hyperparameters}. We also tabulate the active spaces and spin configurations ($S^{2}$, $M=(N_\uparrow-N_\downarrow)/2$ (eigenvalue of $\mathcal S_{z}$)) used to generate the CASSCF baselines. These are listed in Supplementary Table~\ref{table:cas}.

\begin{table*}[htbp]
\caption{Hyperparameters used in calculations. (See ref.~\onlinecite{Hermann2020} for more details.)}\label{table:hyperparameters}
\begin{ruledtabular}
\begin{tabular}{@{\extracolsep\fill}lclc}
 \T Hyperparameter & Value & Hyperparameter & Value\B\\
  \hline
  One-electron basis & 6-311G & Maximum number of determinants & 10\T\\
  Dimension of $\mathbf{e}$ (\# distance features) & 32 & Dimension of $\mathbf{x}_{i}$ (embedding dimension) & 128\\
  Dimension of $\mathbf{z}_{i}$ (kernel dimension) & 64 & Number of layers in $\mathbf{w}_{\boldsymbol\theta}$ & 1\\
  Number of layers in $\mathbf{h}_{\boldsymbol\theta}$ & 2 & Number of layers in $\mathbf{g}_{\boldsymbol\theta}$ & 2\\  
  Number of interaction layers L: &  & Number of layers in $\eta_{\boldsymbol\theta}$: & \\ 
  ~~small/intermediate systems \& ethylene & 4 & ~~small/intermediate systems \& ethylene & 3 \\ 
  ~~benzene & 5 & ~~benzene & 5\\ 
  Number of layers in $\kappa_{\boldsymbol\theta}$: & & Number of walkers: & \\ 
  ~~small/intermediate systems \& ethylene & 3 & ~~small/intermediate systems \& ethylene & 1000\\ 
  ~~benzene & 5 & ~~benzene & 400\\ 
  Batch size & 2000 & Number of equilibration steps & 500\\ 
  Number of training steps: & & Optimizer & AdamW\\  
  ~~small systems & 5000/10000 & Learning rate scheduler & CyclicLR\\ 
  ~~intermediate systems & 20000 & Minimum/maximum learning rate: & \\  
  ~~ethylene & 10000 & ~~small/intermediate systems \& ethylene & 0.0001/0.005\\   
  ~~benzene & 10000 & ~~benzene & 0.0001/0.001\\   
  Cyclic frequency & 1000 & Clipping window q & 5\\   
  Minimum/maximum $\alpha$ & 0.25/1.0 & Epoch size & 10\\     
  Resampling frequency & 3 & Number of decorrelation sampling steps: & \\    
  Target acceptance & 57\% & ~~small/intermediate systems \& ethylene & 5\\ 
   & & ~~benzene & 15\B\\
\end{tabular}
\end{ruledtabular}
\end{table*}

\begin{table*}[htbp]
\caption{The active spaces, specifying $N$ electrons across $P$ orbitals, and spin configurations, used to generate the CASSCF baselines for all systems.}\label{table:cas}
\begin{ruledtabular}
\begin{tabular}{@{\extracolsep\fill}lcclcc}
 \T System & CAS$(P, N)$ & Spin$(S^2, M)$ & System & CAS$(P, N)$ & Spin$(S^2, M)$ \B\\
  \hline
  Li & $(5,3)$ & ($\tfrac34$/$\tfrac34$/$\tfrac34$/$\tfrac34$/$\tfrac34$, 1) &  Be (many states) & $(6,4)$ & (0/2/2/2/0/0/0/2, 1)\T\\
  Be & $(5,4)$ & (0/2/2/2/0, 0) & BH & $(5,2)$ & (0/2/2/0, 0)\\
  B & $(5,5)$  & ($\tfrac34$/$\tfrac34$/$\tfrac34$/$\tfrac{15}4$/$\tfrac{15}4$, 1) & CH$^{+}$ & $(5,2)$ & (0/2/2/0, 0)\\
  C & $(5,6)$ & (2/2/2/0/0, 0) & H$_{2}$O & $(5,2)$ & (0/2/0, 0)\\
  N & $(5,7)$ & ($\tfrac{15}4$/$\tfrac34$/$\tfrac34$/$\tfrac34$/$\tfrac34$, 1) & NH$_{3}$ & $(7,2)$ & (0/2/0, 0)\\
  O & $(5,8)$ & (2/2/2/0/0, 0)& CO & $(10,10)$ & (0/2/0, 0)\\
  LiH & $(12,4)$ & (0/2/0/2/2/0/0/2, 0) & C$_{2}$H$_{4}$ & $(4,4)$ & (0/0, 0)\\
  BeH & $(12,5)$  & ($\tfrac34$/$\tfrac34$/$\tfrac34$/$\tfrac34$/$\tfrac{15}4$/$\tfrac{15}4$/$\tfrac34$/$\tfrac34$, 1) & C$_{6}$H$_{6}$ & $(4,4)$ & (0/2, 0)\B\\
\end{tabular}
\end{ruledtabular}
\end{table*}

\end{document}